\documentclass[10pt,a4paper,twoside]{article}
\usepackage{epsfig}
\usepackage{baltlat6}
\usepackage{array}
\usepackage{here}
\usepackage{amssymb,amsmath}
\begin{document}
\ \
\vspace{0.5mm}
\setcounter{page}{231}

\titlehead{Baltic Astronomy, vol.\,24, 231--241, 2015}

\titleb{Structural and morphological properties of ultraluminous infrared galaxies at $1<z<3$}

\begin{authorl}
\authorb{Guanwen Fang}{1,2},
\authorb{Zhongyang Ma}{2,3},
\authorb{Yang Chen}{2,3,4} and
\authorb{Xu Kong}{2,3}
\end{authorl}


\begin{addressl}
\addressb{1}{Institute for Astronomy and History of Science and Technology, Dali University, Dali 671003, China; wen@mail.ustc.edu.cn}
\addressb{2}{Key Laboratory for Research in Galaxies and Cosmology, CAS, Hefei 230026, China}
\addressb{3}{Center for Astrophysics, University of Science and Technology of China, Hefei 230026, China; xkong@ustc.edu.cn}
\addressb{4}{SISSA, via Bonomea 265, I-34136 Trieste, Italy}
\end{addressl}

\submitb{Received: 2015 January 31; accepted: 2015 May 14}

\begin{summary}
Using the Hubble Space Telescope (HST)/Wide Field Camera 3 (WFC3) near-infrared high-resolution imaging from the 3D-HST survey, we analyze the morphology and structure of 502 ultraluminous infrared galaxies (ULIRGs; $L_{\rm IR}>10^{12}L_{\odot}$) at $1<z<3$. Their rest-frame optical morphologies show that high-redshift ULIRGs are a mixture of mergers or interacting systems, irregular galaxies, disks, and ellipticals. Most of ULIRGs in our sample can be roughly divided into merging systems and late-type galaxies (Sb$-$Ir), with relatively high $M_{20}$ ($>-1.7$) and small S\'{e}rsic index ($n<2.5$), while others are elliptical-like (E/S0/Sa) morphologies with lower $M_{20}$ ($<-1.7$) and larger $n$ ($>2.5$). The morphological diversities of ULIRGs suggest that there are different formation processes for these galaxies. Merger processes between galaxies and disk instabilities play an important role in the formation and evolution of ULIRGs at high redshift. In the meantime, we also find that the evolution of the size ($r_{\rm e}$) with redshift of ULIRGs at redshift $z\sim1-3$ follows $r_{\rm e}\propto(1+z)^{-(0.96\pm0.23)}$.
\end{summary}


\resthead{Morphology and structure of ULIRGs}
{Fang et al.}


\sectionb{1}{INTRODUCTION}

ULtraluminous InfraRed Galaxies (ULIRGs; $L_{\rm 8-1000~\mu m}>10^{12}{~\rm L_{\odot}}$)
were first hinted at by the deep InfraRed Astronomical Satellite (IRAS; Neugebauer et al. 1984) surveys.
Within the past decade, observations have shown that high-redshift ULIRGs are massive
galaxies ($M_{\ast}>10^{10}~\rm M_{\odot}$), with extremely high ratio of infrared to optical
flux density ($F({\rm 24~\mu m})/F(R)>1000$) and intensive
star formation (100--1000 $\rm M_{\odot}~{yr}^{-1}$) (Chapman et al. 2003; Houck et al. 2005; Yan et al. 2007;
Dey et al. 2008; Desai et al. 2009; Huang et al. 2009; Fang et al. 2014).
At high redshift there are many pre-selected ULIRGs samples, such as Dusty-Obscured Galaxies
(DOGs with $(R-[24])_{\rm Vega}>24$; Houck et al. 2005),
SubMillimeter Galaxies (SMGs with $F({\rm 850~\mu m})>0.5{~\rm mJy}$; Chapman et al. 2003),
and Multiband Imaging Photometer for Spitzer (MIPS) 24~$\mu$m selected
samples (Yan et al. 2007), and follow-up analysis is then necessary to single out ULIRGs.

Since the discovery, ULIRGs have been suggested to be a feasible evolutionary phase
towards the formation of local massive early-type galaxies (Sanders et al. 1988;
Veilleux et al. 2009; Hou et al. 2011). But, the existence of a large number of massive galaxies
with $M_{\ast}>10^{10}~\rm M_{\odot}$ at $z\sim2-3$ challenges the merge
theory which massive galaxies assemble at a later time through the merge of smaller
galaxies (Narayanan et al. 2009). During the gas-rich major merger, intense star formation
is triggered and the dust-enshrouded galaxies can be identified as ULIRGs (Wu et al. 1998). At the same time,
it is possible that the gas can be fed into the central massive black holes as quasars.
There are many structural features of mergers, such as multiple bright nuclei, tadpoles
(appear to have undergone a merger by evidence of tails),
irregular shapes, pairs of galaxies depending on the merge stages or the types of merger.
Therefore, morphological and structural studies of $z\sim2$ ULIRGs with or without merger features,
it can help to understand the formation and evolution of massive galaxies
(Shen et al. 2003; Sandage 2005; Ball et al. 2008; Kong et al. 2006, 2009; Fang et al. 2009, 2012).

For galaxies at $1<z<3$, Hubble Space Telescope (HST)/Wide Field Camera 3 (WFC3)
near-infrared (NIR) imaging can provide crucial clues to the rest-frame optical morphologies
($\lambda_{\rm rest}\sim5000$~{\AA}). At such redshift, HST/WFC3 NIR bands have not yet reached
the Balmer break (${\lambda}_{\rm rest}\geqslant 4000~\rm {\AA}$) and probe redder wavelengths
. This will enable us
to study the rest-frame optical morphologies and structures of ULIRGs at $1<z<3$.
By using HST NIR images (NICMOS or WFC3), many groups (Dasyra et al. 2008; Melbourne et al. 2008, 2009;
Bussmann et al. 2009, 2011; Zamojski et al. 2011; Kartaltepe et al. 2012) found the morphologies of ULIRGs
are diverse, e.g., disks, bulges , multiple components, and irregulars. This implies that ULIRGs may
have different formation processes such as mergers and secular evolution without mergers.

Since the samples of previous research programs are commonly small ($<80$ for ULIRGs at $1<z<3$),
it still remains many uncertainties on the structural properties of ULIRGs.
This paper constructs a sample of 502 ULIRGs from the 3D-HST survey\footnote{http://3dhst.research.yale.edu/Home.html}
(Brammer et al. 2012; Skelton et al. 2014). Moreover,
comparing with previous studies based on HST/NICMOS F160W images ($0''.09~{\rm pixel}^{-1}$),
this work will utilize HST/WFC3 NIR images ($0''.06~{\rm pixel}^{-1}$) to investigate the morphological
diversities of high-redshift ULIRGs, and for the first time we explore the size evolution with redshift of our sample
and calculate nonparametric morphological parameters of ULIRGs at $1<z<3$.
Section 2 describes the selection of ULIRGs and the data (include images and catalogs) from the 3D-HST fields.
We present the structural and morphological properties of ULIRGs in Section 3 and 4, and summarize our results in Section 5.
Throughout this paper, we adopt a standard cosmology
$H_{\rm 0}=70$~km s$^{-1}$ Mpc$^{-1}$, $\Omega_\Lambda =0.7$, and $\Omega_{\rm M} = 0.3$.
All magnitudes use the AB system unless otherwise noted.

\sectionb{2}{SAMPLE SELECTION AND DATA}

Total infrared luminosity ($L_{\rm IR}=L_{\rm 8-1000~\mu m}$) is an important measurement in
characterizing ULIRGs at $1<z<3$. Direct measurement of $L_{\rm IR}$ requires far-infrared
photometric data, yet they are not available for most of $24~\mu$m selected sources. This
is particularly true for our sample. In our work, we adopt a luminosity-independent conversion
from the observed Spitzer/MIPS $24~\mu$m flux density to $L_{\rm IR}$, based on a single template
that is the logarithm mean of Wuyts et al. (2008) templates with
$1\leq \alpha \leq 2.5$\footnote{http://www.mpe.mpg.de/$\sim$swuyts/Lir$\_$template.html}.
Wuyts et al. (2011) demonstrated that this luminosity-independent conversion from $24~\mu$m
photometry to $L_{\rm IR}$ yields estimates that are in good median agreement with measurement
from Herschel/Photoconductor Array Camera and Spectrometer (PACS) photometry. Finally, we
construct a sample of 502 ULIRGs with $L_{\rm IR}>10^{12}L_{\odot}$ at redshift $1<z<3$
from the three 3D-HST fields (AEGIS, COSMOS, and GOODS-N), using the photometric data of
Spitzer/MIPS $24~\mu$m from Fang et al. (2014), Muzzin et al. (2013), and Kajisawa et al. (2011),
respectively.

3D-HST is a NIR spectroscopic survey with the HST, designed to study the physical processes that
shape galaxies in the distant universe. The survey contains a great diversity of objects from
high-redshift quasars to brown dwarf stars, but is optimally designed for the study of galaxy
formation over $1<z<3.5$ (Brammer et al. 2012; Skelton et al. 2014).
In addition, it also includes NIR (F125W and F160W) high-resolution
($0''.06~{\rm pixel}^{-1}$) imaging data from the WFC3 on the HST (Grogin et al. 2011;
Koekemoer et al. 2011). The $5~\sigma$ point-source detection limit is brighter
than 27.0 mag in the F160W ($H$) and F125W ($J$) filters. Our study is performed using
the latest data (version 4.1) release of the 3D-HST survey.
The stellar mass ($M_{\ast}$) and photometric redshift ($z$, if there is no
spectroscopic redshift available) we adopt in our work also come from the 3D-HST photometric
catalogs (AEGIS, COSMOS, and GOODS-N). Further details are in
Brammer et al. (2012) and Skelton et al. (2014) for the survey and observational design
and the data products. Figure 1 shows the distributions of $L_{\rm IR}$ and
$M_{\ast}$ of ULIRGs with $1<z<3$ in our sample, and all of them have $M_{\ast}>10^{9.5}~\rm M_{\odot}$.

\begin{figure*}
\centering
\includegraphics[angle=0,width=0.8\textwidth]{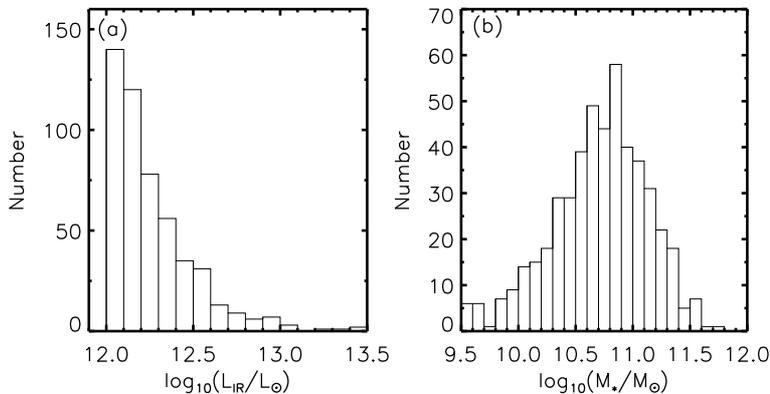}
\caption{(a) Distribution of infrared luminosity ($L_{\rm IR}$) of 502 ULIRGs at redshift $1<z<3$
from the three 3D-HST fields (AEGIS, COSMOS, and GOODS-N). (b) Distribution of
stellar mass ($M_{\ast}$) to ULIRGs in our sample.
} \label{fig:lm}
\end{figure*}

\sectionb{3}{STRUCTURES OF ULIRGS}

Since the redshift distribution of ULIRGs is quite broad ($1<z<3$), we analyze their
rest-frame optical structures on WFC3 F125W or F160W bands according to their redshifts.
For ULIRGs with $1 < z < 1.8$, we choose WFC3 F125W bandpass for structural analysis,
it corresponds approximately to $V$-band in the rest-frame in this redshift range,
but in the redshift range of $1.8<z<3$, we analyze galaxy structure in the rest-frame
optical band ($V$) from the F160W image instead. Finally, 98 ULIRGs in our sample have
$J$-band counterparts ($1 < z < 1.8$), and 404 ULIRGs ($1.8 < z < 3$) are detected in $H$-band image.
The structural parameters of ULIRGs, S\'{e}rsic index ($n$) and effective radius ($r_{\rm e}$),
from the latest catalog\footnote{http://www.mpia-hd.mpg.de/homes/vdwel/candels.html}
(version 1.0) are provided by van der Wel et al. (2012). As described above, we use the observed $J$ structures
at $1<z<1.8$ and the $H$ structures at $1.8<z<3$ for our structural analysis.

Figure 2 shows the $n$ and $r_{\rm e}$ distributions of ULIRGs at different redshift bins
in our sample. From Figure 2(a), the derived S\'{e}rsic indexes ranging from 0.4 to 8,
indicated that a wide range of structural diversities for these ULIRGs, from spheroid
to diffuse structures, e.g., irregulars in appearance, disk-like systems, and elliptical structures.
In total, there are 80\% ULIRGs distribute at $n<2.5$ and 20\% at $n>2.5$.
In addition, we also find the the distribution of sizes of ULIRGs are broad, ranging from 0.5 to 8 kpc,
but most (81\%) of them distribute at $r_{\rm e}<4$ kpc. In Figure 3, the sizes of our ULIRGs sample are
compared to those of $z\sim0.1$ late-type galaxies (LTGs) from Shen et al. (2003). We find that
ULIRGs with $M_{\ast}>10^{10.5}~\rm M_{\odot}$ at $1<z<3$ follow a clear $r_{\rm e}-M_{\ast}$ relation.
However, most of them have smaller sizes, compared to local LTGs with similar stellar
mass. In the meantime, there is also the existence of compact ULIRGs with $r_{\rm e}<1 {\rm kpc}$,
even in massive systems.

\begin{figure*}
\centering
\includegraphics[angle=0,width=0.8\textwidth]{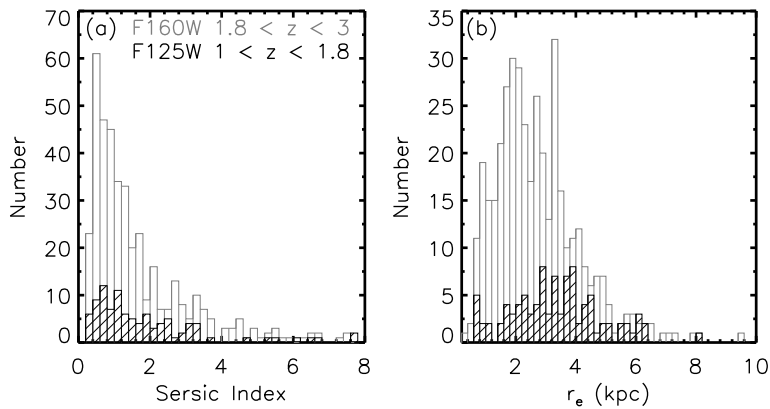}
\caption{S\'{e}rsic index ($n$) and effective radius ($r_{\rm e}$)
histogram of ULIRGs at different redshift bins in our sample.
The left panel (a) is the distribution for $n$,
and the right panel (b) is the distribution for $r_{\rm e}$.
} \label{fig:nre}
\end{figure*}

\begin{figure*}
\centering
\includegraphics[angle=0,width=0.8\textwidth]{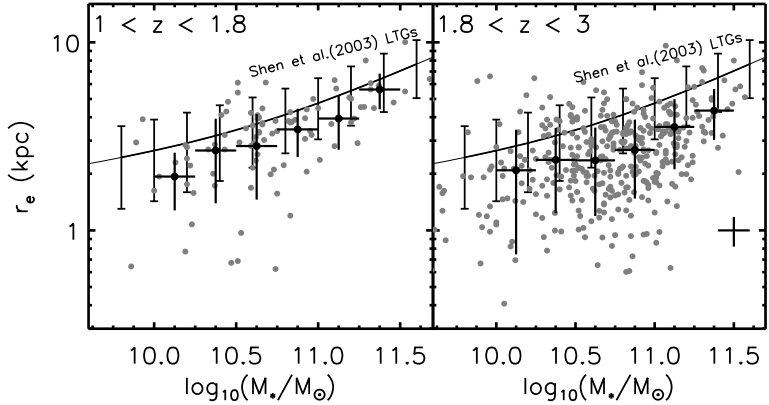}
\caption{Relation of stellar mass ($M_{\ast}$) and effective radius ($r_{\rm e}$) for
ULIRGs at different redshift bins in our sample. The solid lines with $1\sigma$ standard
error are provided by Shen et al. (2003) for local late-type galaxies (LTGs).
Black solid circles represent the median sizes of ULIRGs at different $M_{\ast}$
bins ($\Delta {\rm log}_{10} (M_{\ast}/M_{\odot})=0.25$).
Typical error bars (black) are shown in the right panel.
} \label{fig:rem}
\end{figure*}

In order to explore the size evolution with redshift for ULIRGs at $1<z<3$, we show the sizes of ULIRGs
from our sample in Figure 4. The solid square, triangles, and star in this figure represent the ULIRGs
from Veilleux et al. (2002), Dasyra et al. (2008), and Kartaltepe et al. (2012), respectively.
Based on HST NICMOS $H$-band imaging of 33 $z\sim2$ ULIRGs from a 24 $\mu$m-selected sample of the Spitzer survey,
Dasyra et al. (2008) found that their effective radii range from 1.4 to 4.9 kpc, with a mean
of $<r_{\rm e}>=2.7$ kpc and a dispersion of $\sigma=0.8$ kpc. Using high-resolution HST/WFC3 NIR imaging from CANDELS-GOODS-South field, Kartaltepe et al. (2012) provided the more detailed morphological study of 52 ULIRGs at $z\sim2$.
The median value of sizes of these ULIRGs is $3.3\pm1.7$~kpc. Utilizing the IRAS 1 Jy sample of
118 ULIRGs, Veilleux et al. (2002) found the mean size of local ULIRGs is $4.8\pm1.37$ kpc at $R$ band.

\begin{figure*}
\centering
\includegraphics[angle=0,width=0.75\textwidth]{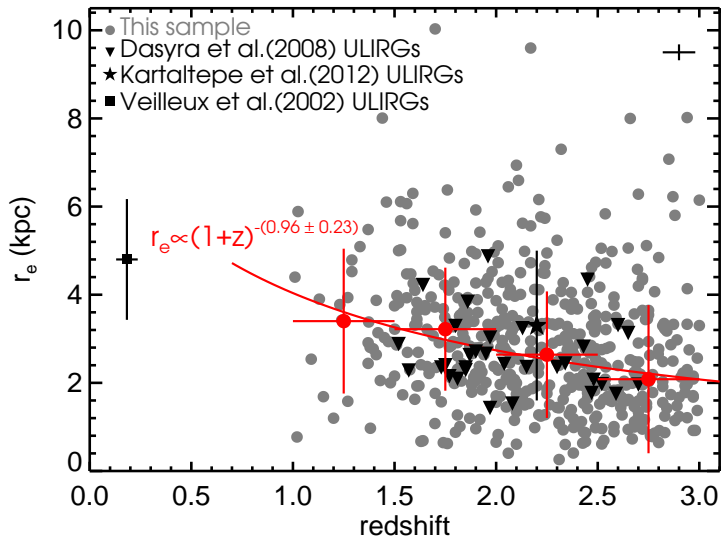}
\caption{Evolution of size with redshift in our ULIRGs sample.
Red solid circles represent the median sizes of ULIRGs at different redshift
bins ($\Delta z=0.5$). Red line corresponds to the best fit for the four
median points ($r_{\rm e}\propto(1+z)^{-(0.96\pm0.23)}$).
The sizes of ULIRGs from the literature are also plotted in this
figure (Veilleux et al. 2002; Dasyra et al. 2008; Kartaltepe et al. 2012).
Typical error bars (black) are shown in the figure.
} \label{fig:rez}
\end{figure*}

In Figure 4, the red solid circles represent the median sizes of our ULIRGs sample at different redshift
bins ($\Delta z=0.5$). The red line, $r_{\rm e}\propto(1+z)^{-(0.96\pm0.23)}$, corresponds to the best fit for the four
median points. The slope ($\alpha=-0.96$) of the size evolution of ULIRGs is steeper than that of gas-rich LTGs
($\alpha=-0.75$ from van der Wel et al. 2014) with similar stellar mass, but it's still far flatter
than the massive early-type galaxies (ETGs) with $\alpha=-1.48$ from van der Wel et al. (2014).
If the Veilleux et al. (2002) data point of local ULIRG $r_{\rm e}$ was included when fitting a power
law to the $r_{\rm e} - z$ relation, we find that the slope ($-0.77\pm0.11$) closer to the LTG value.
A possible explanation is that ULIRGs represent a marginally more compact sub-sample of the LTG population.
This interpretation supported by a large part of our sample is LTGs (see Section 4). Moreover, we find the
sizes of ULIRGs at high redshifts are on average one to two times smaller than those
of local ULIRGs (from Veilleux et al. 2002) with similar infrared luminosity.

\sectionb{4}{MORPHOLOGIES OF ULIRGS}

Morphologies of galaxies correlate a series of physical properties, such as stellar mass,
star formation rate and rest-frame color of galaxies, they can provide direct information
on the formation and evolution history of these objects. Following the method we performed in
Section 3, we use the observed $J$ morphologies at $1 < z < 1.8$ and the $H$ morphologies
at $1.8 < z < 3$ for our morphological analysis. Figure 5 ($J$ band) and
Figure 6 ($H$ band) show examples of the NIR images for ULIRGs
in the COSMOS field of the 3D-HST survey. We perform the visual inspection by three of us,
and find that galaxies in our sample exhibit very diverse morphologies, covering a wide range of types from
interacting systems to compact spheroids. As illustrated in Figure 7,
some of the ULIRGs show morphological features of early-phase mergers,
advanced-phase mergers, or merger remnants. Meanwhile, there are many extended disks and
irregular morphologies for high-redshift ULIRGs.

\begin{figure*}
\centering
\includegraphics[angle=0,width=0.75\textwidth]{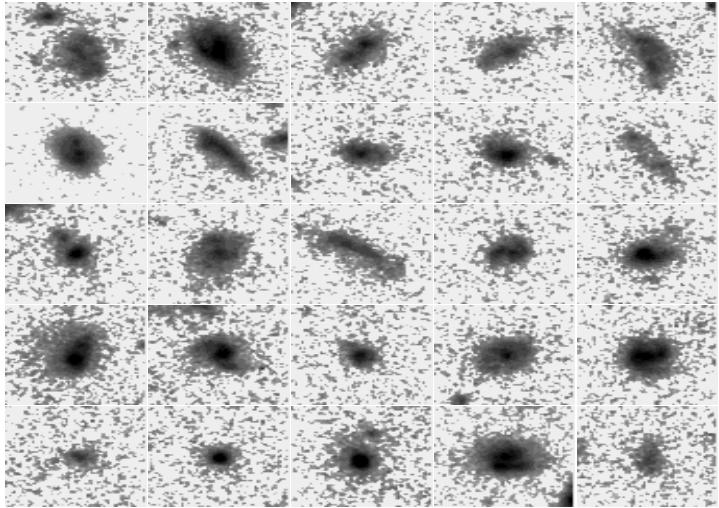}
\caption{HST/WFC3 $J$-band images of ULIRGs at $1<z<1.8$ from the COSMOS field of
the 3D-HST survey. The size of each image is  4$''$$\times$4$''$.
} \label{fig:morJ}
\end{figure*}

\begin{figure*}
\centering
\includegraphics[angle=0,width=0.75\textwidth]{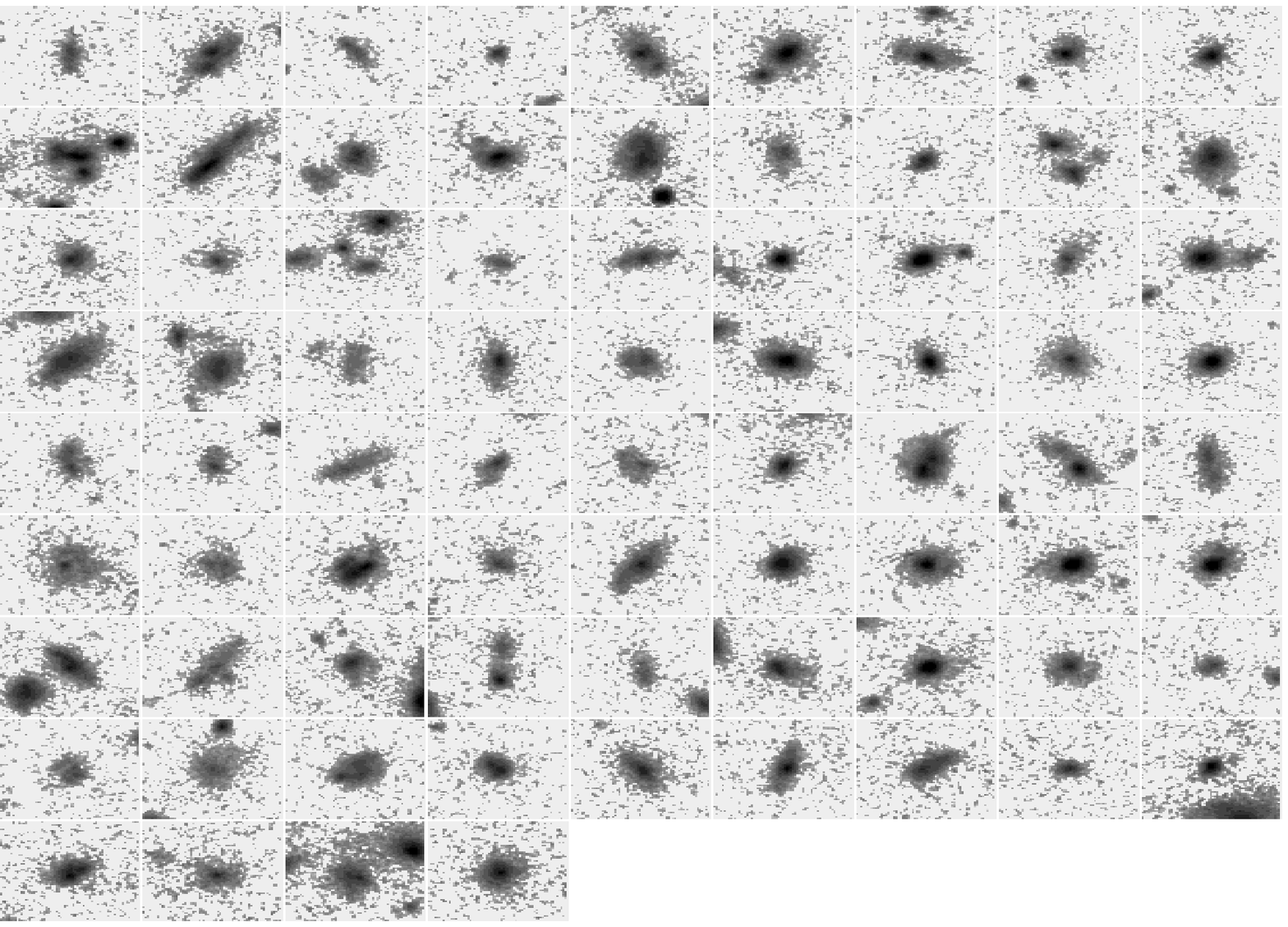}
\caption{HST/WFC3 $H$-band images of ULIRGs at $1.8<z<3$ from the COSMOS field of
the 3D-HST survey. The size of each image is  4$''$$\times$4$''$.
} \label{fig:morH}
\end{figure*}

\begin{figure*}
\centering
\includegraphics[angle=0,width=0.75\textwidth]{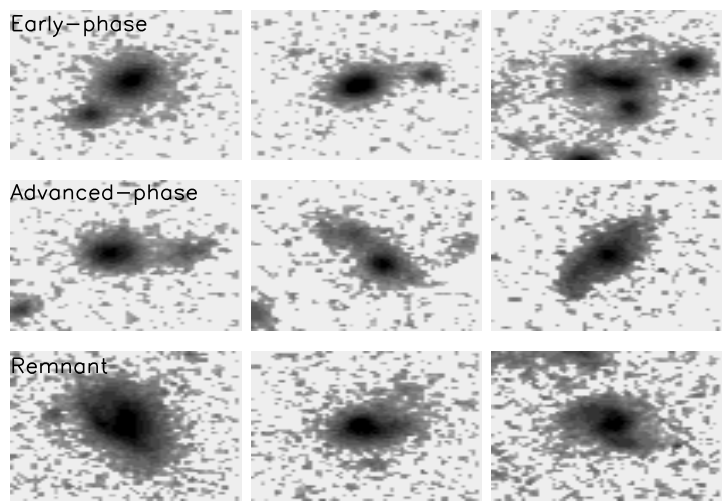}
\caption{Examples of different merging types: early-phase mergers,
advanced-phase mergers, and merger remnants. The size of each image is  4$''$$\times$4$''$.
} \label{fig:merger}
\end{figure*}

In order to quantitatively investigate the morphological features of ULIRGs at $1<z<3$,
we also measure nonparametric morphological parameters (Abraham et al. 1996; Lotz et al. 2004),
such as Gini coefficient ($G$; the relative distribution of the galaxy pixel flux values)
and high moment ($M_{\rm 20}$; the second-order moment of the brightest 20$\%$ of the
galaxy's flux). Based on the rest-frame optical morphologies of galaxies, Lotz et al. (2008)
defined $G$-$M_{20}$ criteria to classify ETGs (E/S0/Sa), LTGs (Sb-Ir), and mergers:

\begin{description}
\item[{\bf ETGs (E/S0/Sa):}] $G\leq-0.14M_{20}+0.33$ and $G>0.14M_{20}+0.80$,
\item[{\bf LTGs (Sb-Ir):}] $G\leq-0.14M_{20}+0.33$ and $G\leq0.14M_{20}+0.80$,
\item[{\bf Mergers:}] $G>-0.14M_{20}+0.33$.
\end{description}

Figure 8 shows the distribution of our sample on the $G$ vs. $M_{20}$ diagram.
For the morphological properties of ULIRGs at $1<z<3$, the majority of them shows mergers and irregular
and disk-like structures, with relatively high $M_{20}$ ($>-1.7$)
and small S\'{e}rsic index ($n<2.5$, $\langle n\rangle=1.4\pm1.3$), while others are elliptical-like (E/S0/Sa)
morphologies with lower $M_{20}$ ($<-1.7$) and larger $n$ ($>2.5$, $\langle n\rangle=3.6\pm1.2$).
Among ULIRGs with $1<z<1.8$ ($1.8<z<3$), the fractions of ETGs, LTGs, and mergers
correspond to 3\% (2\%), 55\% (73\%), and 42\% (25\%), respectively.
This is in agreement with the result of visual morphologies of ULIRGs.
The existence of so many massive galaxies with
stellar masses $M_{\ast}\gtrsim10^{10}$ at high redshifts challenges the
merger scenario for the formation of massive galaxies. Current
numerical simulations (Narayanan et al. 2009) have failed
to produce as many major mergers as required to explain the
observed number of ULIRGs at $1<z<3$. An alternative formation scenario for ULIRGs:
a massive, gas-rich galaxy could have a SFR as high as $180-500~M_{\odot}~{\rm yr^{-1}}$
without any merging process. The diversity
of morphologies indicates that ULIRGs may occur in different
interaction stages of major mergers, in minor mergers, or via
secular evolution not involving mergers at all.

For ULIRGs in our sample, the fraction of objects classified as ETGs
only is small, and remains roughly constant across the full
luminosity/redshift range. The fraction of galaxies classified as LTGs decreases dramatically with luminosity
while the fraction of mergers and interactions increases. The fraction of mergers and interactions
among the $1.8<z<3$ ULIRGs is lower than at $1<z<1.8$ while
the fraction of LTGs is higher at the similar IR luminosity and the same rest-frame wavelength.
This suggests that there has been a evolution in the morphology of ULIRGs between these two redshifts.

Star-forming galaxies in the local universe follow a tight
correlation between stellar mass and star formation rate (SFR), defining a main
sequence (MS; Brinchmann et al. 2004). The
MS is also seen at $0.5 < z < 3$ (Noeske et al. 2007; Elbaz et al.
2007; Daddi et al. 2007). Galaxies with SFGs elevated significantly above ($2\times{\rm MS}$)
this relation are considered to be starbursts.
For ULIRGs in our sample, about 65\% of objects have significantly elevated SFRs relative to the normal MS.
This implies that violent starburst play an important role in ULIRGs at $z\sim2$.

\begin{figure*}
\centering
\includegraphics[angle=0,width=0.8\textwidth]{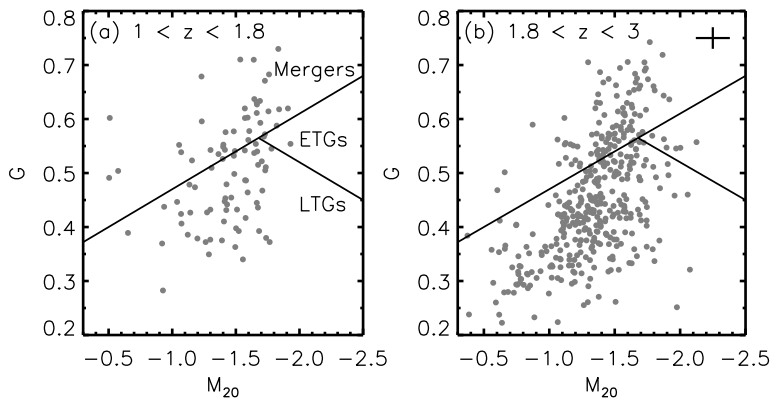}
\caption{Distribution of the rest-frame optical ($\sim5000$~{\AA}) morphologies of ULIRGs
in the $M_{\rm 20}$ vs. Gini coefficient plane. The solid lines represent the
defined criteria of Lotz et al. (2008). Early-type galaxies (ETGs, E/S0/Sa): $G\leq-0.14M_{20}+0.33$ and
$G>0.14M_{20}+0.80$. Late-type galaxies (LTGs, Sb$-$Ir): $G\leq-0.14M_{20}+0.33$ and $G\leq0.14M_{20}+0.80$.
Mergers: $G>-0.14M_{20}+0.33$. Typical error bars (black) are shown in the right panel.
} \label{fig:gm20}
\end{figure*}

\sectionb{5}{SUMMARY}

In this paper, we construct a sample of 502 ULIRGs with $L_{\rm IR}>10^{12}L_{\odot}$ at $1<z<3$
from the 3D-HST survey (AEGIS, COSMOS, and GOODS-N). Utilizing HST/WFC3 NIR (F125W and F160W)
high-resolution images, we study the morphological and structural diversities of these galaxies
in the rest-frame optical. To clearly depict the morphologies of ULIRGs at $z\sim2$, we perform nonparametric measures of
galaxy morphology. In the meantime, we explore the size ($r_{\rm e}$) evolution with redshift of our sample.

We find the rest-frame optical morphologies of high-redshift
ULIRGs are a mixture of mergers or interacting systems, irregular galaxies,
disks, and ellipticals. Most of ULIRGs in our sample can be roughly divided into
merging systems and late-type galaxies (LTGs), with relatively high $M_{20}$ ($>-1.7$)
and small S\'{e}rsic index ($n<2.5$), while others are elliptical-like
morphologies with lower $M_{20}$ ($<-1.7$) and larger $n$ ($>2.5$). The morphological
diversities of ULIRGs suggest that there are different formation processes for these galaxies.
Merger processes between galaxies and disk instabilities play an important role in the formation
and evolution of ULIRGs at high redshift.

For the structural properties of ULIRGs in our sample, we find that
ULIRGs at $1<z<3$ follow a clear $r_{\rm e}-M_{\ast}$ relation.
However, most of them have smaller sizes, compared to local LTGs with similar stellar
mass. Meanwhile, we also find that the evolution of
the size with redshift of ULIRGs at $z\sim1-3$ follows $r_{\rm e}\propto(1+z)^{-(0.96\pm0.23)}$.
The slope ($\alpha=-0.96$) of the size evolution of ULIRGs is steeper than that of gas-rich LTGs
($\alpha=-0.75$) with similar stellar mass, but it's still far flatter
than the massive early-type galaxies (ETGs) with $\alpha=-1.48$,
suggesting that ULIRGs represent a marginally more compact sub-sample of the LTG population.
Moreover, we also find the sizes of ULIRGs
at high redshifts are on average one to two times smaller than those of local ULIRGs with similar
infrared luminosity.

\thanks{This work is based on observations taken by the 3D-HST Treasury Program (GO 12177 and 12328) with the NASA/ESA HST, which is operated by the Association of Universities for Research in Astronomy, Inc., under NASA contract NAS5-26555. This work is supported by the National Natural Science Foundation of China (NSFC, Nos. 11303002, 11225315, 1320101002, 11433005, and 11421303), the Specialized Research Fund for the Doctoral Program of Higher Education (SRFDP, No. 20123402110037), the Strategic Priority Research Program ``The Emergence of Cosmological Structures" of the Chinese Academy of Sciences (No. XDB09000000), the Chinese National 973 Fundamental Science Programs (973 program) (2015CB857004), the Yunnan Applied Basic Research Projects (2014FB155) and the Open Research Program of Key Laboratory for Research in Galaxies and Cosmology, CAS.}

\newcommand{\apj} {ApJ}
\newcommand{\apjl} {ApJL}
\newcommand{\apjs} {ApJS}
\newcommand{\aap} {A\&A}
\newcommand{\mnras} {MNRAS}
\newcommand{\araa} {ARA\&A}
\newcommand{\nat} {NATURE}
\newcommand{\pasj} {PASJ}
\newcommand{\aj} {AJ}
\newcommand{\aaps} {A\&AS}

\References

\refb Abraham, R. G., Tanvir, N. R., Santiago B. X. et al.\ 1996, \mnras, 279, L47

\refb Ball, N.~M., Loveday, J.,
\& Brunner, R.~J.\ 2008, \mnras, 383, 907

\refb Brammer, G.~B., van
Dokkum, P.~G., Franx, M., et al.\ 2012, \apjs, 200, 13

\refb Brinchmann, J., Charlot, S., White, S. D. M., et al.\ 2004, \mnras, 351, 1151

\refb Bussmann, R.~S., Dey,
A., Lotz, J., et al.\ 2009, \apj, 693, 750

\refb Bussmann, R.~S., Dey,
A., Lotz, J., et al.\ 2011, \apj, 733, 21

\refb Chapman, S.~C., Blain,
A.~W., Ivison, R.~J., et al.\ 2003, \nat, 422, 695

\refb Daddi, E., Dickinson, M., Morrison, G., et al.\ 2007, \apj, 670, 156

\refb Dasyra, K.~M., Yan, L.,
Helou, G., et al.\ 2008, \apj, 680, 232

\refb Desai, V., Soifer, B.~T.,
Dey, A., et al.\ 2009, \apj, 700, 1190

\refb Dey, A., Soifer, B.~T.,
Desai, V., et al.\ 2008, \apj, 677, 943

\refb Elbaz, D., Daddi, E., Le Borgne, D., et al.\ 2007, \aap, 468, 33

\refb Fang, G.-W., Kong, X., \& Wang, M.\ 2009,
RAA, 9, 59

\refb Fang, G., Kong, X., Chen,
Y., et al.\ 2012, \apj, 751, 109

\refb Fang, G., Huang, J.-S.,
Willner, S.~P., et al.\ 2014, \apj, 781, 63

\refb Grogin, N.~A., Kocevski,
D.~D., Faber, S.~M., et al.\ 2011, \apjs, 197, 35

\refb Hou, L.~G., Han, J.~L.,
Kong, M.~Z., et al.\ 2011, \apj, 732, 72

\refb Houck, J.~R., Soifer,
B.~T., Weedman, D., et al.\ 2005, \apjl, 622, L105

\refb Huang, J.-S., Faber,
S.~M., Daddi, E., et al.\ 2009, \apj, 700, 183

\refb Kajisawa, M.,
Ichikawa, T., Tanaka, I., et al.\ 2011, \pasj, 63, 379

\refb Kartaltepe, J.~S.,
Dickinson, M., Alexander, D.~M., et al.\ 2012, \apj, 757, 23

\refb Koekemoer, A.~M.,
Faber, S.~M., Ferguson, H.~C., et al.\ 2011, \apjs, 197, 36

\refb Kong, X., Daddi, E., Arimoto, N., et al.\ 2006, \apj, 638, 72

\refb Kong, X., Fang, G.-W., Arimoto, N., et al.\ 2009, \apj, 702, 1458

\refb Lotz, J. M., Primack, J., \& Madau, P.\ 2004, \aj, 128, 163

\refb Lotz, J.~M., Davis, M.,
Faber, S.~M., et al.\ 2008, \apj, 672, 177

\refb Melbourne, J., Desai,
V., Armus, L., et al.\ 2008, \aj, 136, 1110

\refb Melbourne, J.,
Bussman, R.~S., Brand, K., et al.\ 2009, \aj, 137, 4854

\refb Muzzin, A., Marchesini,
D., Stefanon, M., et al.\ 2013, \apjs, 206, 8

\refb Narayanan, D., Cox,
T.~J., Hayward, C.~C., et al.\ 2009, \mnras, 400, 1919

\refb Neugebauer, G.,
Habing, H.~J., van Duinen, R., et al.\ 1984, \apjl, 278, L1

\refb Noeske, K. G., Weiner, B. J., Faber, S. M., et al.\ 2007, \apj, 660, L43

\refb Sandage, A.\ 2005, \araa, 43, 581

\refb Sanders, D.~B., Soifer,
B.~T., Elias, J.~H., et al.\ 1988, \apj, 325, 74

\refb Shen, S., Mo, H.~J.,
White, S.~D.~M., et al.\ 2003, \mnras, 343, 978

\refb Skelton, R.~E.,
Whitaker, K.~E., Momcheva, I.~G., et al.\ 2014, \apjs, 214, 24

\refb van der Wel, A., Bell, E.~F., H{\"a}ussler, B., et al.\ 2012, \apjs, 203, 24

\refb van der Wel, A.,
Franx, M., van Dokkum, P.~G., et al.\ 2014, \apj, 788, 28

\refb Veilleux, S., Kim,
D.-C., \& Sanders, D.~B.\ 2002, \apjs, 143, 315

\refb Veilleux, S., Rupke,
D.~S.~N., Kim, D.-C., et al.\ 2009, \apjs, 182, 628

\refb Wu, H., Zou, Z.~L., Xia, X.~Y., et al.\ 1998, \aaps, 132, 181

\refb Wuyts, S., Labb{\'e}, I.,
Schreiber, N.~M.~F., et al.\ 2008, \apj, 682, 985

\refb Wuyts, S., F{\"o}rster
Schreiber, N.~M., Lutz, D., et al.\ 2011, \apj, 738, 106

\refb Yan, L., Sajina, A., Fadda,
D., et al.\ 2007, \apj, 658, 778

\refb Zamojski, M., Yan, L.,
Dasyra, K., et al.\ 2011, \apj, 730, 125

\end{document}